\documentstyle[eqsecnum,aps,prd,multicol,psfig]{revtex} 

\begin{document}




\title{Microscopic Universality and the Chiral Phase Transition
in two Flavor QCD}

\author{F. Farchioni}
\address{Deutsches Elektronen-Synchrotron DESY, Notkestr. 85, D-22603 Hamburg,
Germany}
\author{\vspace{-1mm}Ph. de Forcrand}
\address{Institut f{\"u}r Theoretische Physik, 
ETH-H\"onggerberg, CH-8093 Z{\"u}rich, Switzerland}
\author{\vspace{-1mm}I. Hip}
\address{NIC - John von Neumann Institute for Computing,
FZ-J{\"u}lich, D-52425 J{\"u}lich, Germany}
\author{\vspace{-1mm}C.~B. Lang}
\address{Institut f{\"u}r Theoretische Physik,
Karl-Franzens-Universit\"at Graz, A-8010 Graz, Austria}
\author{\vspace{-1mm}K. Splittorff}   
\address{The Niels Bohr Institute, Blegdamsvej 17, DK-2100 Copenhagen \O, Denmark} 
\date{\today} 
\maketitle

\begin{abstract}
We re-analyze data from available finite-temperature QCD  simulations near the
chiral transition, with the help of Chiral Random Matrix Theory (chRMT). 
Statistical properties of the lowest-lying eigenvalues  of the staggered Dirac
operator for SU(3) lattice gauge theory with dynamical fermions are examined.
We consider temperatures below, near, and above the critical temperature $T_c$
for the chiral phase transition. Below and above $T_c$ the statistics are in
agreement with the exact  analytical predictions in the microscopic scaling
regime. Above $T_c$ we observe a gap in the spectral density and a distribution
compatible with the Airy distribution. Near $T_c$ the eigenvalue correlations
appear inconsistent with chRMT. 
\end{abstract}

\begin{multicols}{2}
\narrowtext

\section{Introduction} 

It is inherent to the lattice approach to quantum field theory, that one has to
extrapolate from finite lattices, finite statistics and non-critical coupling
parameters to infinite lattices, infinite statistics and critical points. Since
the result supposedly is a non-trivial, non-perturbatively defined quantum
field theory, this process is plagued by uncertainties. A typical example of
such a situation, where all these aspects combine, is the study of the thermal
transition in QCD for small quark masses.  One is interested in the continuum
limit (gauge coupling $g\to 0$), small or vanishing fermion masses ($m\to 0$),
close to critical temperature ($T\to T_c$) in the thermodynamic limit ($L_x\to
\infty$) --- a formidable problem.

The extrapolations are always based on assumptions on the asymptotic 
behavior.  Well known examples are scaling functions based on renormalization
group and chiral perturbation theory --- an expansion around a ground state
with  Goldstone bosons. Here we will examine another such approach, which
should allow the extrapolation to infinite volume and vanishing fermion mass:
Chiral Random Matrix Theory (chRMT). 

\subsection{Chiral Random Matrix Theory}

RMT attempts to identify universal features of ensembles of (random) matrices
with common symmetry properties. Its chiral version, if successful, allows to
separate two aspects of a theory like QCD: the general universal properties
shared with other theories from the model-specific ``dynamical'' content of the
theory. Microscopic eigenvalue distribution shapes are an example for the first
aspect,  expectation values of the fermion condensate for the second.

The limitations for validity of the chRMT\space  considerations (for a given
$T$ in the phase of broken chiral symmetry) are set by \cite{LeSm92,ShVe93}
\begin{equation}\label{QCDrange}
\frac{1}{\Lambda_{QCD}} \ll L_x \ll \frac{1}{m_\pi} \; ,
\end{equation}
where $L_x$ is the linear size of the system and $m_\pi$ is the mass of the
lightest (pseudo-)Goldstone boson. The second restriction imposes that the pion
does not  fit into the space-time volume and it therefore appears to be
unphysical. However,  various correlators in the Dirac operator spectrum can be
computed precisely in this limit.

chRMT\space has been proven to give exact analytical predictions for the
spectrum of the Dirac operator in the microscopic limit \cite{AkDaMa97}. The
microscopic scaling region is simply a blowup of the origin in the spectrum. To
be specific, one considers eigenvalues $\lambda$ on the scale $\frac{\pi}{V
\Sigma}$ where $\Sigma$ is the chiral condensate, related to the spectral
density per unit volume $\rho(\lambda)$ via the Banks-Casher \cite{BaCa80}
relation, $\Sigma = \lim_{\lambda\to 0}\lim_{V\to\infty} \pi \rho(\lambda)$.
This regime is, by definition, only well defined in the spontaneously broken
phase where $\rho(0)\not=0$. In the phase with restored symmetry the scale of
interest is set by the density of states in the vicinity of the onset of
$\rho(\lambda)$.

Here we present a study of the microscopic correlators in the spectrum of the
staggered Dirac operator in SU(3) gauge theory with dynamical fermions at
finite temperature. Specifically we examine the low lying eigenvalue statistics
at temperatures below, near, and above the critical temperature of the chiral
phase transition. 

Our analysis is based on the evaluation of the MILC collaboration's gauge
configurations \cite{MILC96,MILCNERSC}.  We therefore concentrate on the new
aspects connected to RMT ideas for the spectral correlators of the Dirac
operator. In particular for $T>T_c$ we study the singularity at the inner
endpoint of the spectral density.

\subsection{Temperature transition}

Strictly speaking, in the continuum limit $a\to 0$, non-zero temperature is
realized for lattices $n_x^3\times n_t$ with $T=1/(n_t\,a(\beta_g))$ and
$n_t/n_x\to 0$. In that limit, for vanishing quark mass $m$, one expects a
phase transition at $T_c$.  In \cite{MILC96} the critical temperature was
estimated to lie between 143 and 154 MeV. For $T<T_c$ chiral symmetry is broken
spontaneously, with massless pseudoscalars and 
$\left\langle\overline{\psi}\psi\right\rangle\neq 0$; above $T_c$ we expect
restoration of this symmetry. Whereas for pure Yang-Mills theory the
deconfinement transition is associated with a breaking of the center-symmetry
with the Polyakov loop as order parameter, this symmetry is explicitly broken
by the fermion action. Nevertheless, remnants of the original breaking feature
of the Polyakov loop persist even for  small fermion masses.

The nature of the chiral phase transition depends on the number of flavors
$N_f$. An argument based on a 3D $\sigma$-models analysis \cite{PiWi84}
predicts a first order phase transition for $N_f\ge 3$.  For $N_f=2$ one  
expects a second order phase transition with $SU(2)\times SU(2)\simeq O(4)$
scaling behavior. However, even first order behavior may be arguable
\cite{Wi92,RaWi93,Ra95}. For staggered fermions at non-vanishing lattice
spacing the correct counting of flavors is unclear since flavor symmetry is
restored only in the continuum limit. Staggered fermions (as simulated by MILC
with the hybrid R-algorithm \cite{GoLiRe87}) correspond to the case $N_f=2$ in
the continuum limit.  On coarse lattices the so defined fermions should show at
least $U(1)\simeq O(2)$ scaling behavior. For a discussion of the various
scenarios cf. \cite{La98,DeTar98,BeDeGo99}.

It is unclear whether the phase transition at $T_c$ extends towards $m>0$ or
whether, when moving from lower to higher temperature, one just observes
crossover-like behavior. Some evidence points towards this second scenario
\cite{AoFuHa98,La98}. In the following, we denote the crossover (phase
transition) position by $T_c(m)$ or simply $T_c$.

\section{Expectations from chRMT}\label{SecRMT}

According to the nature of the elements in the random matrix, chRMT\space
appears in three universality classes. In this paper we consider the SU(3)
gauge theory with quarks in the fundamental representation, which belongs to
the universality class of the chiral unitary ensemble (chUE). The partition
function under study is \cite{ShVe93,JaVe96,WeScWe96}
\begin{eqnarray}
{\mathcal Z}^{(N_f)}(\{m\})&=&
\int {\rm d}M \,e^{-N{\rm Tr} U(M^2)}\nonumber\\
\label{Z}
 &&\times \prod_{f=1}^{N_f}\det(M+T+im_f) \;,
\end{eqnarray}
where $M$ is a $2N\times2N$ block hermitian matrix (the elements of $W$
being random complex numbers), and
$T$ is a deterministic, i.e. not-random, off-diagonal block matrix
\begin{equation} \label{Tmodel}
M=\left(\begin{array}{cc} 0 & W \\ W^\dagger & 0 \end{array} \right) \qquad , 
\qquad   
T=\left(\begin{array}{cc} 0 & t \\ t & 0 \end{array} \right) \; .
\end{equation}
Here d$M$ denotes the Haar measure, $U(x)$ is an analytic function.

The predictions from chRMT\space concern the correlations between the
eigenvalues $\lambda$ of $D\equiv M+T$ on the scale of individual  eigenvalues
in the thermodynamic limit $N\to\infty$.  The matrix $D$  is the analogue of
the massless Dirac operator in QCD.  The chiral phase transition within
chRMT\space is identified through the value of the spectral density of the
eigenvalues of $D$ near zero,  i.e. using the Banks-Casher relation.  

Modeling the chiral phase transition in chRMT\space amounts to driving a
depletion of eigenvalues of $D$ near the origin by means of some temperature
parameter.  Two separate approaches have been examined in the literature.
First, the unitary invariant chRMT\space \cite{ShVe93,AkDaMa97}, corresponding
to (\ref{Z})  with $T=0$, in which the chiral phase transition is driven by
tuning $U(M^2)$. Second, the non-unitary invariant chRMT\space
\cite{JaVe96,WeScWe96,SeWeGu98}, corresponding to (\ref{Z})  with $U(M^2)=M^2$,
where the deterministic block matrix $T$ mimics the effect of the temperature.
In this paper we do not need  to distinguish between the two approaches as they
are consistent for the quantities measured here.

Below $T_c$,  i.e. when $\rho(0)\not=0$, chRMT\space predicts \cite{NiDaWe98} 
the probability distribution for the smallest eigenvalue (for the trivial 
topological sector) 
\begin{equation}\label{Pmin}
P(z,\{\mu\})=\frac{z}{2}\,e^{-z^2/4}\,\frac{\det_{1\leq i,j\leq
    N_f}C_{ij}(\{\sqrt{\mu^2+z^2}\})}{\det_{1\leq i,j\leq
    N_f}A_{ij}(\{\mu\})} \ , 
\end{equation}
with 
\begin{eqnarray}
A_{ij}(\{\mu\})&\equiv&\mu_i^{j-1}I_{j-1}(\mu_i)\;, \nonumber\\
C_{ij}(\{\mu\})&\equiv&\mu_i^{j-1}I_{j+1}(\mu_i)\;, \nonumber\\
z&=&2\,\pi\,\lambda\,\rho(0)\,N\;, \;\textrm{and}\;\;
\mu_f=2\,\pi\,m_f\,\rho(0)\,N\;.
\end{eqnarray}
where $\rho(0)$ is the spectral density at the origin for the {\em massless}
situation (i.e. when $m_f=0$ in (\ref{Z})). $I_j$ denotes the $j$'th modified
Bessel function. This result is universal in the chRMT\space context,  that
is, the analytic form of $P(z,\{\mu\})$ does not change under deformations of
$U(M^2)$ provided that $\rho(0)\not=0$.  After the identification $V\equiv
2\,N$ ($V$ is the physical volume in lattice units),  (\ref{Pmin}) allows to
extract  $\Sigma=\pi\,\rho(0)$ (the fermion condensate in the chiral limit) 
from  finite-volume Dirac spectra. Of course, in this the mild condition
(\ref{QCDrange}) must be satisfied.

Above $T_c$, when there is a finite gap in  the spectral density, the
repulsion between the eigenvalue pair $\pm\lambda_{\rm min}$ becomes
negligible; chRMT\space hence predicts a {\em soft} inner edge, known as the
Airy-solution  \cite{Fo93,TrWi94}.

At $T_c$ --- signaled in chRMT\space by a power-like behavior of  the spectral
density at small $\lambda$ --- $\rho(\lambda) \propto
\lambda^{\frac{1}{\delta}}$, the prediction from chRMT\space is not unique. It
turns out that the spectral correlators depend on the value of $\delta$
\cite{AkDaMa98}.  

The distribution of the smallest eigenvalue is a spectral one-point
correlation function and is quite sensitive to statistical fluctuations (see
below). As an additional measure we also study a two-point correlator: the
level spacing distribution $P(s)$.  Note that the level spacings, 
$s=s_{i+1}-s_i$, are determined in the unfolded spectrum $\{s_i\}_{i=1}^{2N}$.
Unfolding separates the fluctuation properties of the spectrum from the
supposedly smooth background behavior. The unfolded variable is defined in
terms of the eigenvalue spectrum and the local average spectral density by 
\begin{equation}
s_i=\int_{0}^{\lambda_i}\langle\rho(\lambda)\rangle{\rm d}\lambda \ .
\label{defUnf}
\end{equation}
The RMT prediction for the level spacing distribution is well approximated by
the unitary Wigner surmise
\begin{equation}
P(s)=\frac{32}{\pi^2}\,s^2\,e^{-\frac{4\,s^2}{\pi}} \ .
\label{WUE}
\end{equation}
The level spacing distribution is not expected to be affected by temperature 
and masses in chRMT, see e.g. \cite{PuRaWe98}.

chRMT\space makes predictions for average spectral correlators in sectors with
definite topological charge $\nu$, i.e. derived assuming exact zero modes
(these are not included in the  predicted  distributions; cf.
\cite{Da99} for the result of the weighted summation of all topological
sectors). For Ginsparg-Wilson fermions \cite{GiWi82},  which realize chiral
symmetry on the lattice, one may identify exact zero modes as resulting from
topological excitations according to the Atiyah-Singer index theorem (for
Wilson fermions one can hypothesize that zero modes are replaced by real
modes). This is not the case for staggered fermions, where exact zero  modes
are absent away from the continuum limit \cite{SmVi87} and even gauge
configurations with non-vanishing topological charge do not give zero
eigenvalues. Exact zero modes are here  replaced by ``almost'' zero modes which
accumulate to the origin  in the continuum limit. In the strong coupling region
the microscopic staggered Dirac spectra summed over all topological sectors
show \cite{DaHeNi99,BeMeWe98} good agreement with the analytical  prediction
for the topologically neutral, $\nu=0$, sector from chRMT.  However,
approaching weaker coupling observations contradicting this scenario have been
found in a two-dimensional context \cite{FaHiLa99}.

Before turning to the numerical studies, let us comment on the validity of the 
chRMT\space predictions. The condition for application of  chRMT\space in
lattice  analyses is well established when $T\sim 0$: The range in the unfolded
spectrum over which the chRMT\space correlations dominate is $|\lambda|\le
\frac{f^2_\pi}{\langle\bar{\psi}\psi\rangle L_x^2}$  \cite{JaNoPa98,OsVe98a},
where $f_\pi$ is the pion decay constant. An equivalent statement for $T\geq
T_c$ is not known and no stringent tests of the low-lying eigenvalue statistics
have been carried out so far. 

Let us emphasize that even though the larger part of the studies of 
chRMT\space have been focused on the situation where  $\rho(0)\not=0$, there
is nothing wrong from first principles in using chRMT\space when $\rho(0)=0$.

\section{Gauge field configurations and analysis}

By courtesy of the MILC collaboration \cite{MILC96} there are  sets of gauge
configurations \cite{MILCNERSC} available to  the lattice community. These were
generated with two species of dynamical staggered fermions, at various lattice
sizes, temperatures, values of the gauge coupling and small values of the bare
fermion mass. In Table~\ref{tabconf} the samples used in the present study are
listed. For further details on the method of determination of the gauge
configurations and the physical parameters we refer to \cite{MILC96}.

\vspace{15pt}
\begin{table}[h]
\caption{ Summary of the MILC configurations used in our analysis (for FT01 we
only considered a subset of the total of 149 configurations available). The
suggestions in the rightmost column are based on MILC's results. The transition
is near $\beta$=5.26 for $n_t=4$ and $\beta$=5.725 for $n_t=12$.}
\begin{tabular}{lrlrlll}
MILC-set &  \# conf's&$n_x$& $n_t$& $\beta_g$& $m\,a(\beta_g)$&phase\\
\tableline
124A   &  61   &   12  &  4  &  5.25  &  0.0125 &$<T_c$ \\ 
124B   &  91   &   12  &  4  &  5.26  &  0.0125 &$<T_c$ \\
124C   &  126  &   12  &  4  &  5.27  &  0.0125 & near $T_c$ \\
124D   &  57   &   12  &  4  &  5.28  &  0.0125 &$>T_c$ \\
1241   &  42   &   12  &  4  &  5.25  &  0.008  &$<T_c$  \\
1242   &  50   &   12  &  4  &  5.255 &  0.008  &$<T_c$ \\
1243   &  45   &   12  &  4  &  5.26  &  0.008  & near $T_c$ \\
1244   &  47   &   12  &  4  &  5.265 &  0.008  & near $T_c$   \\
1245   &  42   &   12  &  4  &  5.27  &  0.008  &$>T_c$ \\
1246   &  40   &   12  &  4  &  5.28  &  0.008  &$>T_c$ \\
FT01   &  30   &   24  & 12  &  5.65  &  0.008  &$<T_c$\\
FT03   &  131  &   24  & 12  &  5.725 &  0.008  & near $T_c$\\
FT04   &  188  &   24  & 12  &  5.8   &  0.008  & $ > T_c$\\
FT05   &  146  &   24  & 12  &  5.85  &  0.008  &$>T_c$\\
\end{tabular}
\label{tabconf}
\end{table}
The massless staggered Dirac operator is anti-hermitian and therefore has
purely  imaginary eigenvalues, lying symmetric to the origin. We determine the
lowest lying eigenvalues with help of the implicitly restarted Arnoldi  method
\cite{So92}, using Chebyshev polynomials to improve the convergence. 

The convergence speed depends on the separation bet\-ween the eigenvalues.  The
configurations below $T_c$ therefore exhibit much slower convergence and for $V
= 24^3\times 12$ the determination is then quite time-consuming. Configurations
that are supposedly above $T_c$ develop a gap for the smallest eigenvalue and
the convergence of the diagonalization is faster. All eigenvalues have been
obtained with a precision of  at least 5 significant digits.

We have determined (on the positive imaginary axis)  the lowest ten eigenvalues
for the $12^3\times 4$ ensembles  and the lowest eight eigenvalues for the
$24^3\times 12$ ensembles.

The  chRMT-prediction for the smallest eigenvalue distribution has been tested 
against quenched QCD \cite{DaHeKr99,EdHeKi99} and dynamical SU(2) 
\cite{BeMeWe98} lattice simulations at $T=0$; there are also recent quenched
QCD results for non-zero temperature \cite{EdHeKi99a}.  RMT distributions for
the bulk nearest-neighbor spacing in both, the confinement and the
deconfinement phase of full QCD have been observed in \cite{PuRaWe98}  on
$6^3\times 4$ lattices  (Wigner surmise, see Eq.~(\ref{WUE})).

\section{Reporting the results } 

We now turn our attention to the measurement of the statistical properties of 
the Dirac operator. Starting with the distribution of all eigenvalues and  then
separating out the lowest and the second lowest eigenvalue we observe the first
indication of a discrepancy with chRMT. In order to investigate this
discrepancy further we study the level spacing distribution. Finally we focus
on the inner edge of the spectrum for $T>T_c$ and measure the exponent for the
singularity of the spectral density.

\subsection{Eigenvalue density}

\begin{figure}[h]
\vspace*{2mm}
\centerline{\psfig{file=figs/fig1a.eps,clip=,width=6.0cm}}
\centerline{\psfig{file=figs/fig1b.eps,clip=,width=8.4cm}}
\centerline{\psfig{file=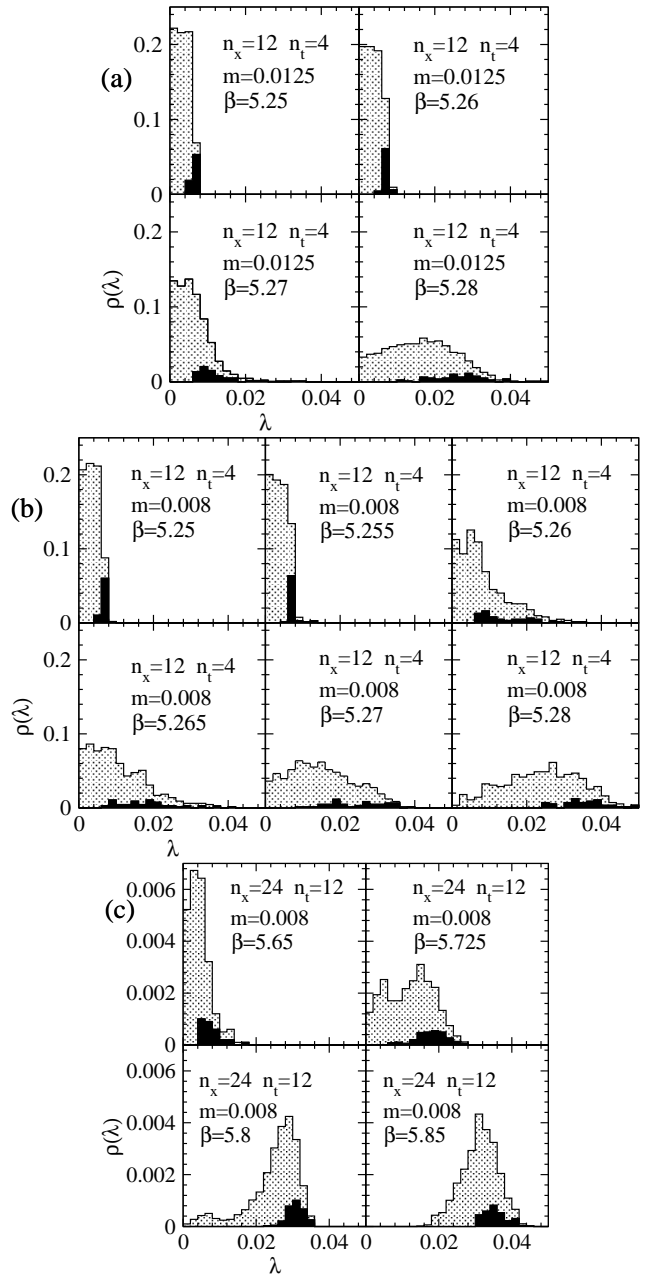,clip=,width=6.0cm}}
\vspace*{2mm}
\caption{Histograms for the 10 (or 8) smallest eigenvalues.
The contribution from the 10th (8th) eigenvalue is indicated in black. 
(a) $12^3\times 4, m=0.0125$, 
(b) $12^3\times 4, m=0.008$,
(c) $24^3\times 12, m=0.008$.}
\label{all-dist}
\end{figure}

Fig.~\ref{all-dist} gives the eigenvalue distribution as obtained from the
lowest 10 (or 8 for the large lattices) eigenvalues for each configuration. It 
coincides with the distribution from all eigenvalues only up to the lowest of
all 10th (8th) eigenvalues. In order to emphasize this feature we also plot the
distribution histogram of this 10th (8th) eigenvalue in full black.

A remarkable change in the features of the  distributions occurs around $T_c$,
as given in Table \ref{tabconf}. In particular for $n_x=24$, $n_t=12$ a gap in
$\rho(\lambda)$ appears for $T>T_c$. 

Since  we do not know about the topological charge of most of  the
configurations studied here, we cannot separate the trivial from the
non-trivial topological sectors, as would be required for a faithful comparison
with chRMT\space distributions.  For sufficiently rough lattices (i.e. in the
\noindent strong coupling region), one can argue \cite{BeMeWe98} that the
topological charge $\nu$ is effectively zero from the fermionic point of view;
however this is no more the case for fine enough lattices \cite{FaHiLa99} and
the problem of the knowledge of the topological charge becomes critical. For
$T>T_c$ the situation is completely different, since topological fluctuations
are suppressed in the continuum theory.

\begin{figure}[h]
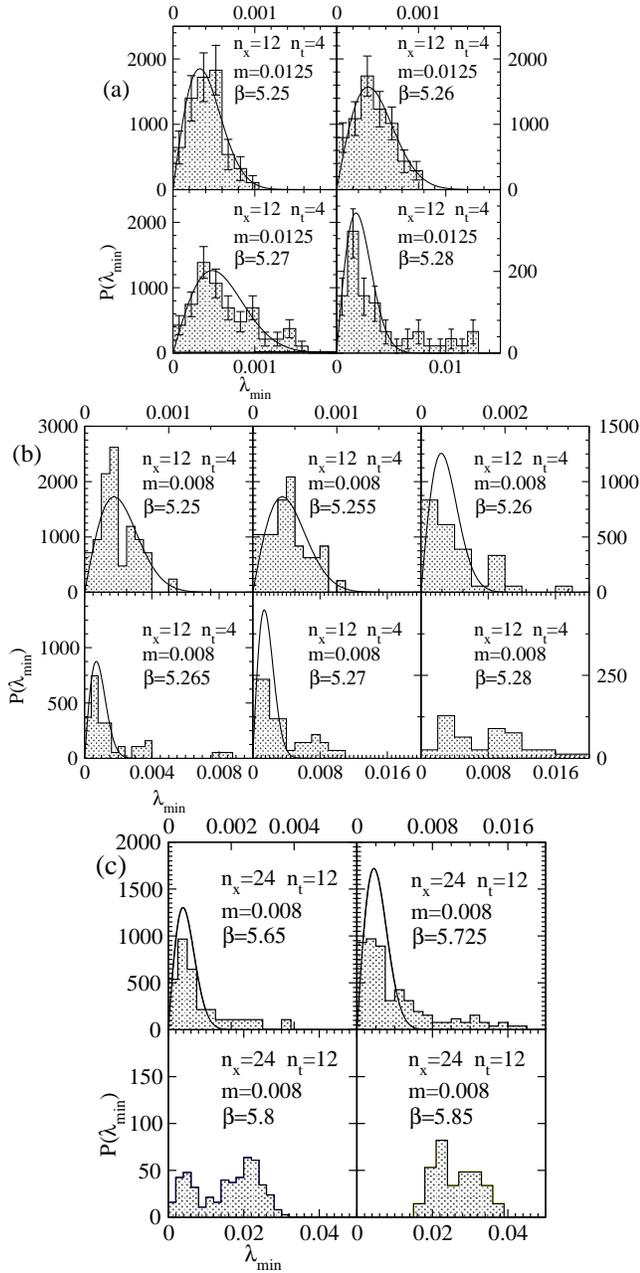

\centerline{\psfig{file=figs/fig2a.eps,clip=,width=6.0cm}}
\centerline{\psfig{file=figs/fig2b.eps,clip=,width=8.4cm}}
\centerline{\psfig{file=figs/fig2c.eps,clip=,width=6.0cm}}
\caption{Histograms for the 1st eigenvalue. (a) $12^3\times
4, m=0.0125$, (b) $12^3\times 4, m=0.008$, (c) $24^3\times 12, m=0.008$. 
For the data where we fitted to chRMT-distribution (cf. Table II) we also plot the
fits. The error bars in (a) are shown to indicate the  typical size of errors in all
histograms. }
\label{firstevs}
\end{figure}

The ``microscopic'' distributions should be in the chUE~ universality class and
prediction (\ref{Pmin}) should apply in particular for the smallest eigenvalue
for $T<T_c$. A fit of the corresponding  prediction  for the topologically
trivial case ($\nu=0$) to the normalized data provides us with the (infinite
volume) parameter $\Sigma$.   We assume the continuum symmetry, i.e. $N_f=2$ in
(\ref{Pmin}). This one-parameter fit appears reasonable only for data
concerning temperatures well below $T_c$ as indicated in
Fig.~\ref{firstevs}(a).  This is made explicit in Table~\ref{tab:fit} where
the  fitted values of $\Sigma$ together with the corresponding $\chi^2/d.o.f.$
are reported; the latter increases with $T$,  and for $T\simeq T_c$ (in
agreement with Table I) prediction (\ref{Pmin}) becomes incompatible with data.

The formal chRMT\space expression (\ref{Pmin}) gives the eigenvalue
distribution as a function of volume and fermion mass; the parameter $\Sigma$ 
(the spectral density at the origin) is thus defined implicitly as the
extrapolation to infinite volume and vanishing fermion mass. From our fits --
if the data follows chRMT formulas -- $\Sigma$ should therefore be independent
of the spatial lattice size $L_x$ and the fermion mass $m$. 

For e.g. $12^3\times 4$, $\beta=5.25$ we find agreement of  $\Sigma$ for
$m=0.0125 $ and 0.008 within the errors. At $\beta=5.26$ the values  disagree.
This value of $\beta$ is, however, close to $T_c$ and the position of the phase
transition (or crossover) is quite sensitive to $m$. Such a change of the
transition point is not accounted for in chRMT, which is not at all sensitive
to the underlying dynamics of QCD.

For comparison we have to extrapolate the MILC  values \cite{MILC96} for the
chiral condensate both, to infinite volume and to vanishing quark mass. Since
(except for chRMT) we have no firm prediction concerning the  volume
dependence, we just extrapolate the MILC results for the large lattices at
$\beta=5.65$ and 5.725 linearly to vanishing quark mass. We know however that
finite-volume effects on the condensate increase as the quark mass decreases.
Therefore, it is not surprising that our linear extrapolation of MILC results
tends to come out slightly but systematically smaller than the values we find
in Table \ref{tab:fit}. 

The issue of the topological sector is likely to be particularly relevant for
the finest lattice at our disposal ($24^3\times 12$), where almost zero modes
could be present and spoil the validity of the trivial sector predictions from
chRMT. These could be the cause of the bump observed for small $\lambda$ at
$\beta=5.8$, both in  the spectral density and in the smallest eigenvalue
distribution. It is therefore with some hesitation, that we compare the
histograms for the smallest  eigenvalues in Fig.~\ref{firstevs} with these
predictions.

\vspace{15pt}
\begin{table}
\caption{Values of the scale $\Sigma$ as obtained from
comparison of lattice data for $P(\lambda_{\min})$ with chUE.}
\begin{tabular}{lrllll}
$n_x$ &$n_t$& $\beta$ & $m$ & $\Sigma$ & $\chi^2/d.o.f.$ \\
\tableline
12 &  4  & 5.25  & 0.0125 & 0.647(29)  & 0.495 \\
12 &  4  & 5.26  & 0.0125 & 0.571(26)  & 0.562 \\
12 &  4  & 5.27  & 0.0125 & 0.449(29)  & 1.651 \\
12 &  4  & 5.28  & 0.0125 & 0.138(13)  & 2.129 \\
12 &  4  & 5.25  & 0.008  & 0.686(37)  & 0.368 \\
12 &  4  & 5.255 & 0.008  & 0.551(52)  & 1.218 \\
12 &  4  & 5.26  & 0.008  & 0.459(43)  & 1.379 \\
12 &  4  & 5.265 & 0.008  & 0.331(39)  & 2.517 \\
12 &  4  & 5.27  & 0.008  & 0.168(21)  & 2.806 \\
24 & 12  & 5.65  & 0.008  & 0.0198(18) & 1.048 \\
24 & 12  & 5.725 & 0.008  & 0.0063(60) & 5.063 \\
\end{tabular}
\label{tab:fit}
\end{table}

In order to further investigate this feature, we studied (for the set  with
$\beta=5. 725$) the influence of the configurations where the eigenvectors
$u_0$ of the lowest eigenvalues exhibit a large contribution to the total
chirality, i.e. $|\langle \bar{u}_0 \gamma_5 u_0\rangle|\geq 0.08$. According
to the index theorem, these configurations with large chirality, which make up
roughly one half of the ensemble, tend to carry non-vanishing topological
charge and therefore zero modes. Indeed  we find that a substantial part (75\%)
of the first peak may be explained from these contributions.

These findings suggest that indeed topological modes are responsible for a
low-lying peak in the distributions. Below $T_c$ all topological sectors are
present and the low-lying eigenvalues have comparable magnitudes (their average
position being roughly proportional to $\nu$).  When the temperature
approaches  $T_c$ the theoretical expectation is  that the topological
fluctuations begin to be suppressed, although still present in the ensemble,
quasi superimposed on the background distribution,  which starts to broaden
significantly with increasing temperature. Sufficiently far above $T_c$ only
the topologically trivial sector survives and there may be no small eigenvalues
at all. This is indeed what we actually observe for the lattice $24^3\times
12$.

In a recent study of quenched configurations \cite{EdHeKi99a} there have been
indications for a dilute gas of  instanton--anti-instanton pairs producing a
Poissonian distribution of small eigenvalues above $T_c$.  These may be
suppressed or absent when considering dynamical fermions. In our context this
seems to be the case for the finest lattices ($24^3\times 12$ with
$\beta=5.85$) at our disposal.

\begin{figure}[h]
\centerline{\psfig{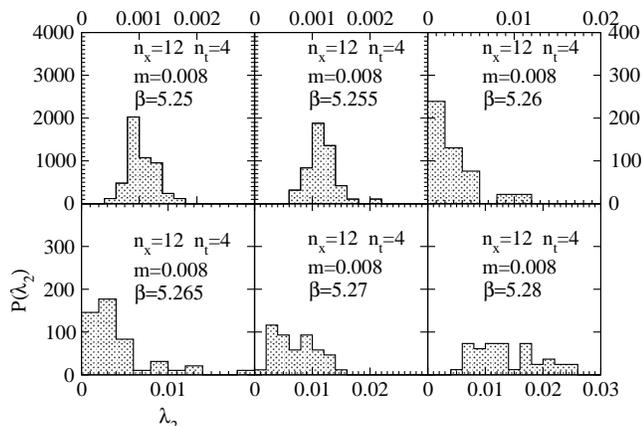}}
\caption{Histograms for the 2nd smallest eigenvalue for
$12^3\times 4, m=0.008$. By comparison with Fig.~\ref{firstevs}(b) we observe that the mutual
overlap between $P(\lambda_{\rm min})$ and $P(\lambda_2)$ increases as
$\beta  \to\beta_c$. This is not consistent with chRMT.}
\label{secondevs}
\end{figure}

Fig.~\ref{secondevs} gives the histograms for the 2nd smallest eigenvalues.
Again we notice a dramatic change of the distribution shape around $T_c(m)$.

We interpret the sudden flatness of the distribution of the smallest
eigenvalues as being (i) due to the vanishing spectral density and (ii) due to
increasing statistical fluctuations near the chiral phase transition. The
latter effect is not reproduced in chRMT\space since this is a zero dimensional
and non-dynamical theory. Furthermore, the mutual overlap of $P(\lambda_1)$ and
$P(\lambda_2)$ increases for $\beta\to\beta_c$.  This is also inconsistent with
chRMT. In order to study this effect further we now turn to the level spacing
distributions.

\subsection{Level spacing distribution}

Another observable with definite predictions from RMT (cf. Sect. \ref{SecRMT})
is the distribution of level spacings. The advantage here is, that the level
spacing should not be influenced by possible distortions of the smallest
eigenvalues due to the unknown topological charge of the configurations (if the
smallest eigenvalues are removed from the data).

The studies of the level spacing statistics in lattice data so far have
shown a uniform picture consistent with the RMT prediction (\ref{WUE}).
The agreement extends on both sides of the confinement-deconfinement phase
transition \cite{PuRaWe98,BeMaPu99}.  However, to the authors' knowledge
there is no analytical prediction from chRMT\space for the level spacing
distribution when focusing on the soft edge or at $T_c$. 
So one might
worry that the standard prediction, Eq.~(\ref{WUE}), is not appropriate
when $\rho(0)=0$.  Within the T-model of \cite{JaVe96} for $T=T_c$ and
$T=3\,T_c$, we have performed a numerical high statistics simulation to
eliminate such doubt, and we there confirmed the distribution (\ref{WUE}).
(The T-model is defined by (\ref{Z}) with $U(M^2)=M^2$ and $t$ in
(\ref{Tmodel}) chosen proportional to the unit matrix.) 

Usually it is  possible to get high statistics on the level spacing
distributions since each  configuration provides a large number of eigenvalue
spacings. However, already the lowest 10 (or 8) eigenvalues allow for a crude
estimate of the  distribution shape. 

Recall that the level spacing distribution is measured in the unfolded
spectrum, see (\ref{defUnf}). Here we use the average spacings $\langle
\lambda_{i+1}-\lambda_i\rangle$ between contiguous eigenvalues to define the
unfolded level spacings 
\begin{equation}
s_{i+1}-s_i={(\lambda_{i+1}-\lambda_i)\over 
\langle \lambda_{i+1}-\lambda_i\rangle }\; .
\label{UnfSpace}
\end{equation}

\begin{figure}[h]
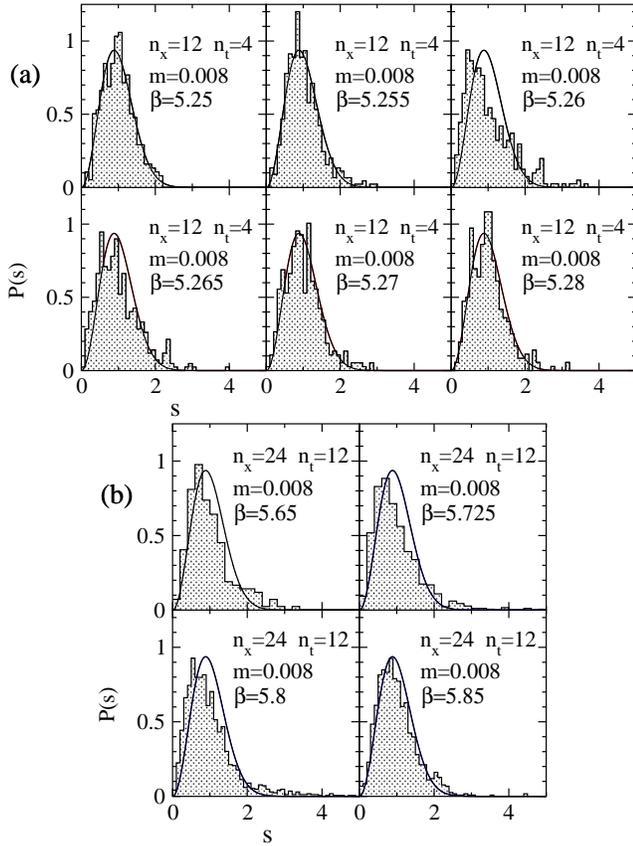

\centerline{\psfig{file=figs/fig4a.eps,angle=270,clip=,width=8.4cm}}
\centerline{\psfig{file=figs/fig4b.eps,angle=270,clip=,width=6.0cm}}
\caption{Histograms for the distribution of (a) the first 9  eigenvalue
spacings on the $12^3\times 4$ lattice and (b) the first 7 eigenvalue spacings
on the $24^3\times 12$  lattice  with $m=0.008$ (according to [5,6] $\beta_c
\simeq 5.7$ in this case). The standard prediction (\ref{WUE}) for the  level
spacing distribution is plotted for comparison.}
\label{levelspc}
\end{figure}

In Fig.~\ref{levelspc} we compare the data with the parameter-free
theoretical expectation. Whereas below and above $T_c$ we find reasonable
agreement with the theoretical expectation, there are clear discrepancies near
$T_c$. We observe  unexpected high histogram entries. Since the average value
by definition is 1 this then leads to a shift of the central peak to the left.

In order to further check our unfolding procedure, we also considered
other approaches, e.g. using a average density as in (\ref{defUnf}) by
smoothing our distribution in various ways. We furthermore tried to
discard the higher lying eigenvalues, e.g. using only the lowest  5 level
spacings or introducing a cutoff near the peak of the distributions in
Fig.~\ref{firstevs}.  In all those checks we found essentially the same
behavior with discrepancies near $T_c$.

In conclusion of this section, we observe at $T\simeq T_c$  a breakdown of the
otherwise universal microscopic spectral correlations. The dynamics of QCD
plays an essential r\^ole in the phase transition. A RMT model where such
dynamics is not there fails to account for the increased  fluctuations in the
eigenvalue level spacings.

\subsection{Soft inner edge}

We now turn to the results for $T>T_c$. In our results for  $24^3\times 12$ at
$\beta=5.85$ (Fig.~\ref{all-dist}(c))  a gap in the spectral distributions  is
obvious. However even at $\beta=5.8$ we may speculate, that a clear signal of a
gap is only prevented by the (topological) quasi-zero modes responsible for the
small bump at small eigenvalues.

Recall, that chRMT\space \cite{JaVe96} predicts the presence of a gap in the
spectral density $\rho(\lambda)$ of the Dirac operator centered around
$\lambda=0$. Furthermore, the inner edge of this gap is predicted to  show a
singularity, at a point $A$, in the macroscopic spectral density \cite{KaFr98}
\begin{equation}
\rho(\lambda,u)=K\left(\frac{\lambda^2}{A^2}-1\right)^{u+1/2} \ ,
\ \ u=0,2,4,...  \  \ ,  
\label{AiSingularity}
\end{equation}
where $K$ is a known constant. The constant $u$ takes the value $u=0$ in the
generic chRMT, i. e. without fine-tuning the matrix potential in (\ref{Z})
(which would be necessary in order to  obtain higher values of $u$). This
corresponds to a square-root-like eigenvalue density near $A$.

One concern here is to measure $u$. With the limited amount of data available
it is not possible to do this based on the spectral density only. Instead we
propose to study the average distance between the smallest and the sequel
eigenvalues $\langle q_{i-\frac{1}{2}}\rangle
\equiv\langle\lambda_{i+1}-\lambda_1\rangle$; the extraction of $u$ is carried
out by noting the following scaling relation  in the index $i$, ordering by
size the eigenvalues which follow the smallest,
\begin{equation}
\langle q_{i-\frac{1}{2}}\rangle\propto \left(i-\frac{1}{2}\right)^\frac{1}{u+3/2} \ .
\label{qi}
\end{equation}
This proportionality follows by integration in (\ref{AiSingularity}). 

\begin{figure}[h]
\centerline{\psfig{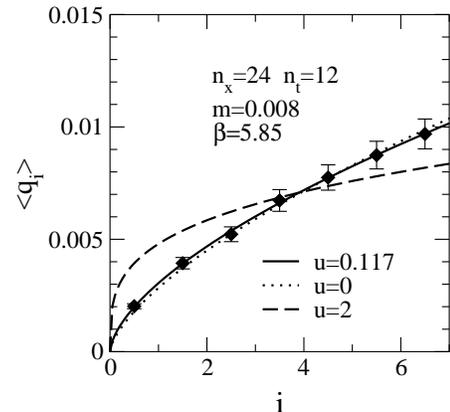}}
\caption{The behavior of the average distance from the smallest eigenvalue to
the $i$'th eigenvalue  for the $24^3\times 12$ ensemble at $\beta=5.85$.}
\label{AND}
\end{figure}

In Fig.~\ref{AND} we display the seven average distances,  $\langle
q_i\rangle$, from the  ensemble of 146 configurations on a $24^3\times12$
lattice for $\beta_g=5.85$. We also exhibit the best fit to (\ref{qi}) with
respect to $u$, giving $u=0.117(71)$. Also shown are the corresponding curves
for $u=0$ and $u=2$. The value $u=0$ is clearly favored.

Since the two-point correlations behave as expected from RMT,
we now turn to the one-point distribution.
The microscopic behavior of the spectral density in the vicinity of this
singularity is universal in the chRMT\space sense, but depends on the value  of
$u$ \cite{KaFr98}. For $u=0$ the exact analytical prediction for the
microscopic spectral density in the vicinity of the inner edge is \cite{Fo93}
\begin{equation}
\rho_{\rm Ai}(z,0)=({\rm Ai}'(z))^2-z\,({\rm Ai}(z))^2 \ .  
\label{AiDens}
\end{equation}
Here the origin has been moved to the inner spectral endpoint $A$ by means
of the rescaled eigenvalue $z$, which is defined through
\begin{equation}
\lambda=A\left[1+\frac{z}{2}\left(\frac{2}{\pi AK}\right)^\frac{1}{u+\frac{3}{2}}\right] \ .
\end{equation}

\begin{figure}[h]
\centerline{\psfig{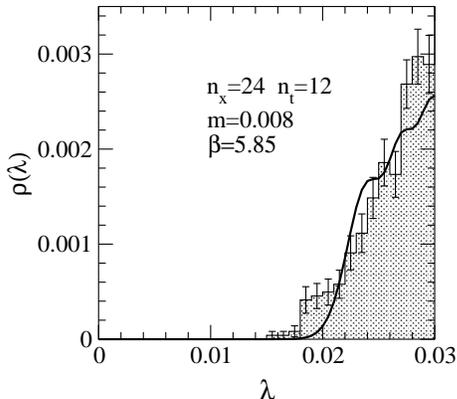}}
\caption{The $u=0$ Airy density with
endpoint $\lambda=A=0.0205$. Approximately the domain of the first three
eigenvalues is shown.}
\label{AiryFig}
\end{figure}

The consistency with the prediction (\ref{qi}) for $u=0$ and the approximate
validity of chRMT\space correlations in the level spacing statistics  above
$T_c$ suggest that the Airy-density (\ref{AiDens}) corresponding to the value
$u=0$ should fit the spectral density. If it does, then we can extract the
inner-endpoint $A$ of the spectrum in the thermodynamic limit, by fitting
$\rho_{\rm Ai}(z,0)$ with respect to $A$ to the lowest part of the spectral
density, see Fig.~\ref{AiryFig}.

This fit does not convincingly confirm Airy density. However,
the statistical fluctuations at this $\beta$-value affect the {\em one-point
distribution} substantially and prevent a decisive comparison.

\section{Summary and conclusions} 

We have examined the manifestations of the chiral phase transition in the
microscopic spectral correlators for the Dirac operator. For the level 
spacing distribution, we find agreement with
RMT\space below and above $T_c$. Below $T_c$ the chRMT\space distributions
allow us to determine condensate values with implicit consideration of lattice
volume and quark mass dependence. This could in principle serve to improve the
scaling analysis of the condensate near the chiral transition.

Near $T_c$, however, the microscopic spectral statistics differs from the
chRMT\space prediction.   By measuring the Monte Carlo time evolution of the
chiral condensate, Aoki et al. \cite{AoFuHa98} have shown, that there are mixed
phase signals, which, however, vanish towards larger volumes. The existence of
a mixed phase would offer an explanation for the observed deviations from 
chRMT\space  near $T_c$. In that case the level spacing distribution near $T_c$
would be a mixture of those from the two phases. Such a mixture would lead to
large spacings: the spacings are unfolded according to the average spacings of
the total ensemble and not according to that of the separate phases.

The observed discrepancy from the RMT level spacing statistics may also be
interpreted as an inclination towards Poissonian statistics; distribution
shapes interpolating between Wigner and Poissonian statistics have been
suggested by Brody \cite{Br73}. 

As may be seen from the $24^3\times 12$ ensembles at $\beta=5.8$ and $5.85$, a
gap develops the spectral density for $T>T_c$. This is consistent with the 
observed  suppression of topological fluctuations in the latter ensemble
\cite{FoGaHe98}. For the $\beta=5.85$ ensemble we have measured the critical
exponent characterizing the steepness of the density at the inner edge. The
value is found to be compatible with 1/2. This is exactly as predicted by
chRMT\space where the chiral phase transition is manifested by the crossover
from the Bessel hard edge to the Airy soft edge. The indications of the Bessel
to Airy scenario are suggestive but simulations with extended statistics are
needed in order to quantify the observation. However, even with low statistics 
the $\beta$-dependence of the distance between e.g. 8th and 1st eigenvalue
provides an excellent means to  identify the change of the phase.

At low $\beta$, on coarse lattices, staggered fermions appear to be blind with
regard to the topological charge of the gauge configurations, and the smallest
eigenvalue distribution agrees with the chRMT\space distribution for the
$\nu=0$ sector. As the lattice becomes finer, topology becomes more relevant.
Although this is maybe ``good'' for the continuum limit of staggered fermions, 
it affects unfavorably the agreement with chRMT\space since the want-to-be-zero
modes and the non-zero modes have similar eigenvalues, and begin to separate
only when the non-zero modes are pushed to larger values when  increasing the
temperature.

\acknowledgements
We want to thank the MILC collaboration for making available the gauge
configurations, that we used in our analysis. Special thanks go to Jim Hetrick
and Doug Toussaint for their help in accessing those and for the support in
making additional ensembles public.  Ph. de~F. thanks Jim Hetrick and
Jean-Fran\c{c}ois Laga\"e for their contribution at a preliminary stage of this
project. K.~S. would like  to thank Andrew Jackson  for discussions.



\end{multicols}
\end{document}